\documentclass{article}
\usepackage[paper=letterpaper,margin=2cm]{geometry}
\usepackage{soul, color, xcolor}
\usepackage{graphicx}
\usepackage{amsmath}
\usepackage{amssymb}
\usepackage{braket}
\usepackage{bm}
\usepackage{amsfonts}
\usepackage{cite}
\usepackage[ruled,boxed,linesnumbered]{algorithm2e}
\usepackage{tcolorbox}
\usepackage[colorlinks=true,citecolor=blue]{hyperref}
\usepackage{booktabs,graphicx}
\usepackage{dashrule}
\usepackage{authblk}
\usepackage{orcidlink}
\usepackage{multirow}
\usepackage{dcolumn}
\tcbuselibrary{breakable}
\tcbset{colbacktitle=yellow!10!white, colback=white, breakable, coltitle=black, before upper={\parindent10pt\noindent}, fonttitle=\bfseries}
\date{}
\title{\textbf{Robust quantum key distribution with imperfections in state preparation and characterization}}

\author{\begin{minipage}{0.92\textwidth}
    \centering
    \small
    Yi-Fei Lu\orcidlink{0000-0001-8212-0556}$^1$
    Yan-Yang Zhou\orcidlink{0009-0006-5630-0717}$^{1,\dagger}$
    Yang Wang\orcidlink{0000-0001-9774-1064}$^1$,
    Xin-Hang Li$^2$,
    Yu Zhou$^1$,
    Xiao-Lei Jiang$^1$,
    Jia-Ji Li$^1$,
    Chun Zhou$^1$,
    Hong-Wei Li$^1$,
    Yu-Yao Guo$^2$,
    Lin-Jie Zhou$^2$,
    Wan-Su Bao$^{1,\ddagger}$

    $^1$Henan Key Laboratory of Quantum Information and Cryptography, IEU, Zhengzhou 450001, China

    $^2$State Key Laboratory of Advanced Optical Communication Systems and Networks, Shanghai Key Lab of Navigation and Location Services, Department of Electronic Engineering, Shanghai Jiao Tong University, Shanghai 200240, China

    $^\dagger$zyy@qiclab.cn, $^\ddagger$bws@qiclab.cn.
\end{minipage}}

\begin{document}
\maketitle

{\centering\section*{Abstract}}

Quantum key distribution (QKD) offers an approach for secure communication over quantum channels, based on the principles of quantum mechanics. However, the inevitable imperfections in QKD systems can introduce critical vulnerabilities, posing significant challenges to the deployment of QKD in terms of both practical security and performance. In this work, we present a comprehensive security analysis of QKD in the practical scenarios involving imperfections in modulation spaces. Specifically, our analysis focuses on the imperfections related to encoding, intensity modulation, and their characterization. We provide a unified framework based on the operator dominance method to overcome these issues and achieve a tight estimation of the secret key rate (SKR). The simulation results show that the SKR in practical scenarios, with parameters characterized from commercial QKD systems, closely matches the SKR under ideal conditions. In the finite case, the maximum transmission distance can reach up to 250 km with total pulse number $N=10^{12}$ and the SKR closely approaches the asymptotic limit up to 250 km with $N=10^{14}$. This validates the robustness of our method against imperfections in state preparation and characterization. Furthermore, the feasibility and robustness of method in measurement-device-independent (MDI) QKD have also been demonstrated. Our analysis not only promotes advancements in enhancing practical security and performance but also highlights the feasibility of deploying QKD for actual applications.

\clearpage

\section{Introduction}

Quantum key distribution (QKD) \cite{bennett2014RN153,xu2020RN130,pirandola2020RN485,azuma2023RN1046} allows information-theoretically secure key agreement between distant parties (named as Alice and Bob) over a quantum channel. QKD has witnessed numerous significant achievements in both theoretical \cite{grosshans2002RN526,long2002RN1066,islam2017RN762,lucamarini2018RN45,zeng2022RN816,xie2022RN724} and experimental \cite{yin2016RN288,boaron2018RN286,chen2021RN519,liu2023RN975,zhou2023RN978,fadri2023RN1023,li2023RN987,zheng2025RN1274} aspects. By employing the decoy-state method \cite{hwang2003RN202,lo2005RN81,wang2005RN185}, QKD can be conveniently implemented using weak coherent light sources. Meanwhile, measurement-device-independent (MDI) QKD \cite{lo2012RN72,braunstein2012RN581,ma2012RN73,tamaki2012RN74} eliminates the quantum hacking attacks targeting the receiver, significantly enhancing the practical security of QKD. However, the issues of source-side imperfections have not been fully addressed, constraining both the practical security and the implementation performance.

In QKD, the source is responsible for preparing a set of quantum states based on the encoding bits. These states are typically divided into two bases: one basis is used for key generation, while the other is employed to estimate information leakage. However, in practical QKD systems, the prepared quantum states are inevitably imperfect, yielding deviated pure states or mixed states \cite{huang2023RN959}. The imperfections may arise from issues in optical devices (e.g., lasers, modulators, optical attenuators), distortions in the electrical driving signals, and the presence of noise. In single-photon QKD, the modulation imperfection corresponds to the encoding space, which is called as flawed source \cite{tamaki2014RN69}. In the more general case of decoy-state QKD with weak coherent source, the state preparation encompasses not only the encoding but also the intensity modulation, where the imperfection of the latter is referred as intensity fluctuation \cite{wang2007RN1145}. Any imperfections in state preparation could potentially be exploited by an attacker, Eve, to compromise the practical security of the QKD system \cite{gottesman2004RN266,2010RN237,tang2013RN305,sajeed2015RN313,li2018RN805}. Moreover, a significant but frequently neglected issue is the reliability of characterizing the actual prepared states. This is because most security analyses are based on full or partial characterization of the prepared quantum states. But, due to inherent limitations of practical devices, the characterization of prepared states is inevitably subject to imperfections such as noise and systematic deviations. It is imperative that the practical security analysis of QKD with state preparation imperfections must also take into account the imperfections in state characterization. Therefore, the vulnerabilities in modulation spaces include imperfections related to encoding, intensity modulation, and the characterization of these parameters.

At present, several security proofs have been proposed to tackle the imperfections of state preparation in various aspects. Firstly, most analyses focus specifically on the issue of imperfect encoding. These methods can be broadly categorized into three classes based on their assumptions about state characterization. The first class requires a precise characterization of the actual prepared states. For example, the original loss-tolerant (LT) method \cite{tamaki2014RN69,xu2015RN402,mizutani2015RN435,tang2016RN433,lorenzo2021RN559,boaron2018RN286,bourassa2020RN648} is proposed, which characterizes the quantum channel through Pauli decomposition and gives a rigorous and tight security analysis. This improvement stems from the fact that the LT method's decomposition depends on the specific form of the prepared states, thereby requiring the accurate characterization. From this perspective, the original LT method assumes full knowledge of the practically prepared quantum states, thereby requiring accurate state characterization. Alternatively, the virtual mutually unbiased bases method \cite{li2024RN1179} can also be employed to address the characterized encoding imperfections. The second class relaxes this requirement by only demanding partial characterization. Based on a bound of the fidelity between the two bases, it could provide a security analysis with a flawed source \cite{gottesmanRN1285,lo2007RN264,koashi2009RN137}. However, as it conservatively assumes Eve can enhance the flaws by exploiting the transmission loss, it provides a pessimistic estimation of the secret key rate (SKR). The improved LT method eliminates the accurate state characterization of the original version. It is generalized as long as intervals of the modulation are characterized \cite{nagamatsu2016RN64,mizutani2019RN61,huang2023RN959,guillermo2023RN1332}. On the one hand, these methods are more robust and easier to implement. On the other hand, since the analysis relies solely on the bounds of the imperfection intervals, it conservatively adopts worst-case assumptions, which inevitably sacrifices a certain degree of performance \cite{huang2023RN959}. The third class goes further by treating the source as uncharacterized, making minimal assumptions about the prepared states. It proposes a security analysis method for MDI-QKD based on uncharacterized encoding states \cite{yin2014RN1161,wang2016RN1159,hwang2017RN1162,zhu2019RN1158,zhou2020RN494}. But it will show a pessimistic estimation of the SKR with large flaws. Recently, the case of basis misalignment is considered to give a tolerant analysis for MDI-QKD \cite{lu2022RN835}. Secondly, the intensity parameters are essential for estimating the counting rate and phase error rate of single-photon states in decoy-state QKD. To address intensity imperfections, the worst-case scenario based on the bounds of the pulse intensities is considered to derive reliable bounds for the single-photon states with analytical formulas or linear programming method \cite{wang2007RN1145,wang2009RN306,li2014RN733,lu2019RN408,lu2021RN1368,jiang2022RN1146,lu2023RN1143}. This has low requirements for intensity characterization, but faces the issue of yielding loose parameter estimation. Thirdly, the collective analysis is necessary when both encoding and intensity imperfections are considered simultaneously. When encoding and intensity imperfections are assumed to be independent, the above analysis methods can be directly combined. For example, the LT method has been generalized to enhance its practicality by adapting it to scenarios where Alice only needs to characterize the intervals of encoding and intensity fluctuations \cite{nagamatsu2016RN64,mizutani2019RN61,huang2023RN959,guillermo2023RN1332}. Overall, we can see a trade-off between the requirement for state characterization and QKD performance. Accurate state characterization can enable higher performance but is experimentally challenging. In contrast, simplified state characterization is easier to implement and more robust, but the resulting conservative estimation based on these state characterizations inevitably comes at the cost of sacrificing a certain degree of performance.

In this work, we aim to address both encoding and intensity imperfections and balance the requirement for state characterization and QKD performance. Our method is based on a characterization of the quantum states (not merely the intervals of encoding and intensity), while allowing for some tolerance in the accuracy of the state characterization. This flexibility helps reduce conservativeness in parameter estimation so as to achieve high QKD performance while maintaining practicality. By presenting the operator dominance method \cite{maeda2019RN507,wang2020RN798} to characterize these imperfections, we establish a unified framework that simultaneously overcomes the imperfections of encoding, intensity, and characterization. The operator dominance method was proposed in Ref. \cite{maeda2019RN507} for the security analysis of twin-field QKD. Here, we use the prepared quantum states to dominate the single-photon states, thereby ensuring a unified and tight estimation of the SKR. And we demonstrate that both encoding and intensity imperfections can be effectively addressed without requiring precise characterization of the quantum states. The comparison of analyses with state preparation and characterization imperfections between our and current methods is summarize in Tab. \ref{tab_1}. To validate our method, we conduct numerical simulations of the SKR with practical parameters characterized from commercial QKD systems \cite{huang2023RN959}. The results show that the SKR under practical conditions closely approximates the ideal case. In the finite case, the transmission distance can reach 250 km with total pulse number $N = 10^{12}$, and the SKR closely approaches the asymptotic limit up to 250 km with reasonable $N = 10^{14}$. Furthermore, we provide a concise analysis and simulation to demonstrate the applicability and tightness of this method to MDI-QKD. Our work guarantees the robustness and effectiveness of practical QKD systems by providing a framework to handle real-world imperfections, which could contribute to the advancement of secure quantum communication.

\begin{table*}[]
\caption{Comparison of security analysis methods with state preparation and characterization imperfections.}
\label{tab_1}
\resizebox{\linewidth}{!}{
\begin{tabular}{@{}ccccc@{}}
\toprule
\multirow{2}{*}{Approach}                           & \multicolumn{2}{c}{\begin{tabular}[c]{@{}c@{}}State   preparation\\     imperfections\end{tabular}} & \multirow{2}{*}{\begin{tabular}[c]{@{}c@{}}State   characterization\\    imperfections\end{tabular}} & \multirow{2}{*}{SKR}                 \\ \cmidrule(lr){2-3}
                                                    & Encoding          Intensity                                         &                                                                                                         &                                      \\ \midrule
Original LT \cite{tamaki2014RN69}                   & $\checkmark$      & $\times$        & $\times$         & Tight                                \\
Improved LT \cite{mizutani2019RN61}                 & $\checkmark$      & $\checkmark$    & $\checkmark$     & \multirow{2}{*}{\begin{tabular}[c]{@{}c@{}}Tight/Loose with small/large\\    characterization uncertainty\end{tabular}}  \\                                 \\
Uncharacterized-source methods \cite{yin2014RN1161} & $\checkmark$      & $\times$        & $\checkmark$     & Tight/Loose with small/large imperfections \\
Present   work                                      & $\checkmark$      & $\checkmark$    & $\checkmark$     & Tight                                \\ \bottomrule
\end{tabular}}
\end{table*}

\section{Imperfections in state preparation and characterization}

In this section, we give an analysis of the imperfections in state preparation and characterization, and give the representation of these imperfections to facilitate the security analysis. 

In perfect single-photon QKD systems, Alice prepares the perfect quantum states $\ket{0},\ket{1}$ in Z basis for key generation and $\ket{+},\ket{-}$ in X basis for the estimation of information leakage. However, the practical prepared states may exhibit deviations from the ideal states, leading to imperfections. The practical states with encoding imperfection can be generally denoted as 
\begin{equation}
    \label{eqj}
    \rho_i \in \mathcal{H}_2, i \in \{0,1,+,-\},
\end{equation}
where $\mathcal{H}_2$ is the two-dimensional Hilbert space. The deviations from the ideal states are characterized by the fidelity $F(\ket{i},\rho_i) = \sqrt{\bra{i} \rho_i \ket{i}}$. When the practical states are pure, they can be expressed as $\rho_i = \hat{P}(\ket{\varphi_i})$, where $\hat{P}(\ket{\cdot}) = \ket{\cdot}\bra{\cdot}$. Here, we note that the prepared states in Eq. (\ref{eqj}) implicitly imply that the states do not include the potential side-channels and corrections, which are referred as the qubit assumption and independent and identically distributed (i.i.d.) assumption \cite{tamaki2014RN69}.

The weak coherent source is widely used in QKD systems due to the inefficiency of the current single-photon source. The coherent state with the intensity $\tau$ and phase $\theta$ can be denoted as $\ket{e^{i\theta} \sqrt{\tau}}$. And the phase-randomized weak coherent state is equivalent to a mixture of the photon number states $\sum_{j=0}^\infty p_{j,\tau} \ket{j}\bra{j}$, where $p_{j,\tau}$ is the probability of Poisson distribution and $\ket{j}$ is the photon number states. The ideal encoding states $\ket{i},i\in\{0,1,+,-\}$ and the photon number states $\ket{j},j\in\mathbb{N}$ share some similarities in their definitions. Below, we do not explicitly distinguish them unless clarification is needed. We note that the secret keys can only be distilled from the single-photon states, while the multi-photon states will introduce the side channels. The decoy-state method addresses this issue, which provides a tight estimation corresponding to the single-photon states. In the decoy-state method, a set of states with different intensities are prepared: one for key generation and others for parameter estimation. Therefore, there are two modulation spaces in decoy-state QKD: the encoding space for key encoding, and the intensity space used for varying the photon number distribution. The intensity modulation may also exhibit imperfections, which are manifested as the intensity imperfection (i.e., intensity fluctuation). The phase-randomized weak coherent state with intensity imperfection can be expressed as
\begin{equation}
    \label{eqr}
    \kappa_\tau = \sum_{j=0}^\infty q_{j,\tau} \ket{j}\bra{j},\tau \in \Pi,
\end{equation}
where $\Pi$ is the set of intensities, $q_{j,\tau}$ may not be the Poisson distribution due to the intensity imperfection. Suppose the encoding and intensity are independent, the prepared state with encoding and intensity imperfections can be shown as $\kappa_\tau \otimes \rho_i$. For example, the single-photon state is $\ket{1}\bra{1} \otimes \rho_i$, where $\ket{1}\bra{1}$ denotes the photon number and $\rho_i$ denotes the encoding $i\in\{0,1,+,-\}$.

In practical QKD system, the actual prepared state $\kappa_\tau \otimes \rho_i$ can be characterized experimentally, and the results show that the imperfections of the encoding and intensities both follow the Gaussian distributions \cite{huang2023RN959}. If the states are characterized perfectly, then we can perform security analysis based on the characterized results. However, due to the limitations in characterization precision and the presence of noise, the performance of actual devices cannot guarantee precise state preparation. We refer this issue as the imperfection in state characterization. If the security analysis is based on the inaccurate states, it may lead to an insecure SKR. Therefore, it is essential to develop methods that account for these imperfections in state characterization. We denote the state characterized under conditions of both state preparation and characterization imperfections as
\begin{equation}
    \kappa_\tau^\prime \otimes \rho_i^\prime, i \in \{0,1,+,-\}, \tau \in \Pi.
\end{equation}
The relationship between this state and the actual state $\rho_i \otimes \kappa_\tau$ satisfies that $F(\kappa_\tau \otimes \rho_i,\kappa_\tau^\prime \otimes \rho_i^\prime) = F(\kappa_\tau,\kappa_\tau^\prime) F(\rho_i ,\rho_i^\prime) \leq 1$. Actually, the characterized states $\rho_i^\prime$ and $\kappa_\tau^\prime$ may not significantly deviate from the actual states $\rho_i$ and $\kappa_\tau$. This means that even with imperfections in state characterization, the characterized states $\kappa_\tau^\prime \otimes \rho_i^\prime$ can still provide a reference in a way for the characterization of the actual states $\kappa_\tau \otimes \rho_i$. We give a representation of this characterization as $\kappa_\tau \otimes \rho_i \in V_\tau \otimes V_i$ or more explicitly, as follows:
\begin{equation}
    \begin{aligned}
        \rho_i &\in V_i \triangleq V(\rho_i^\prime), \\
        \kappa_\tau &\in V_\tau \triangleq V(\kappa_\tau^\prime).
    \end{aligned}
\end{equation}
Here, $V(\rho_i^\prime)$ and $V(\kappa_\tau^\prime)$ represent the minimal spaces that are guaranteed to contain the actual quantum states $\rho_i$ and $\kappa_\tau$, based on the states $\rho_i^\prime$ and $\kappa_\tau^\prime$ characterized in practice. Based on this representation, we can present the security analysis of QKD with the imperfections in state preparation and characterization.

\section{Security analysis of QKD with imperfections}

In single-photon QKD, only encoding is required, while both encoding and intensity modulation are necessary in decoy-state QKD. Therefore, we begin by providing the security analysis of encoding and characterization imperfections for single-photon QKD. This foundational analysis is directly applicable to decoy-state QKD. Subsequently, by incorporating the imperfections in intensity modulation and characterization, we achieve a comprehensive security analysis of the imperfections in modulation spaces for QKD.

\subsection{Encoding and characterization imperfections}

Suppose Alice's encoding states are $\rho_i$ ($i\in\{0,1,+,-\}$) in Eq. (\ref{eqj}), which will be sent to Bob for detection. Since these states are ultimately held and measured by Bob, we denote the corresponding system as $b$. We can purify the states in system $b$ by introducing the ancillary system $A$ as
\begin{equation}
    \ket{\phi_i}_{Ab} = \sum_{m\in\{0,1\}} \ket{\gamma_{i,m}}_A \ket{m}_b.
\end{equation}
Here, the (unnormalized) ancillary states in the system $A$ are virtual and we restrict only the correlations such that the inner products satisfy $\braket{\gamma_{i,n} | \gamma_{i,m}} = \braket{m | \rho_i | n}$. This ancillary system is assumed to be possessed by Alice locally \cite{tamaki2014RN69,bourassa2020RN648}. The states $\ket{m}$ with $m\in\{0,1\}$ correspond to the Z basis in system $b$. If the states are characterized with deviations as $\rho_i^\prime$, we still could assume the virtual purification similarly and obtain the states $\ket{\phi_i^\prime}_{Ab}$. This is because the purification is entirely virtual and does not affect the actual states and their measurement results.

We consider an entanglement protocol where Alice prepares a composite state in Z basis by introducing another ancillary system $a$ to record the bits as
\begin{equation}
    \label{eqo}
    \ket{\Phi_z}_{aAb} = \sum_{i\in\{0,1\}} \sqrt{p_i} \ket{i}_a \ket{\phi_i}_{Ab}.
\end{equation}
If Alice measures the system $a$ in the basis $\{\ket{0},\ket{1}\}$, she will obtain the state $\ket{i}_a \ket{\phi_i}_{Ab}$ with probability $p_i$. The states in the system $b$ are still $\rho_i = \text{Tr}_{aA} [\hat{P}(\ket{i}_a \ket{\phi_i}_{Ab})]$, showing the equivalence of the entanglement scheme and the prepare-and-measure scheme. We note that the system $A$ for the purification of system $b$ does not affect the practical process. Since the operations on the system $a$ are commuted with Eve and Bob's operations on the system $b$, Alice could postpone the measurement on system $a$ after Bob receives the states. After Bob receives the states, the composite state in systems $a$ and $b$ possessed by Alice and Bob can be expressed as
\begin{equation}
    \begin{aligned}
        \mathcal{E}(\rho_{ab}) = \text{Tr}_{A} \big\{ \mathcal{E} [\hat{P}(\ket{\Phi_z}_{aAb})] \big\}
        = \sum_{i,m,j,n \in\{0,1\}} \sqrt{p_i p_j} \braket{\gamma_{j,n} | \gamma_{i,m}} \ket{i}\bra{j}_a \otimes \mathcal{E} (\ket{m}\bra{n}_b),
    \end{aligned}
\end{equation}
where $\mathcal{E}$ denote Eve's evolution on the system $b$. Then Alice and Bob measure the systems $a$ and $b$ in Z basis to obtain the raw key bits, respectively. Bob's measurement in Z basis can be defined as the positive-operator valued measure (POVM) $\{\hat{M}_0,\hat{M}_1,\hat{M}_\bot\}$, where $\hat{M}_{0(1)}$ and $\hat{M}_\bot$ corresponds to the result $0(1)$ and inconclusive outcome, respectively. The SKR of the single-photon QKD is given by \cite{bennett2014RN153,lo1999RN138,shor2000RN300}
\begin{equation}
 R_s = p_z^2 s_1 [1 - h(e_x) - h(e_z)],
\end{equation}
where $p_z$ is the probability of Z basis, $s_1$, $e_x$, and $e_z$ are the counting rate, phase error rate, and the bit error rate of the single-photon states, respectively. And $h(x) = -x\log_2(x) - (1-x) \log_2(1-x)$ is the binary Shannon entropy function. 

The parameters $s_1$ and $e_z$ can be obtained directly with the raw key bits. However, the phase error rate should be estimated to characterize the information leakage. Suppose that Alice and Bob have measured the systems $a$ and $b$ of the state $\mathcal{E}(\rho_{ab})$ in X basis, the phase error rate can be defined as \cite{koashi2007RN398,koashi2009RN137,tamaki2014RN69}
\begin{equation}
    \label{eqa}
 e_x = \frac{Y_{+,-}^{(z)} + Y_{-,+}^{(z)}}{Y_{+,+}^{(z)} + Y_{+,-}^{(z)} + Y_{-,+}^{(z)} + Y_{-,-}^{(z)}},
\end{equation}
where $Y_{i,j}^{(z)}$, with $i,j = \pm$, denotes the joint probability that Alice and Bob obtain results $i$ and $j$ when both measuring the composite state $\mathcal{E}(\rho_{ab})$ in X basis. As the system $a$ is local and virtual, the measurement on the system $a$ can be supposed perfect. When Alice's measurement results on the system $a$ is $\pm$, the state in the system $b$ can be shown as 
\begin{equation}
    \label{eqd}
    \varrho_\pm = \sum_{m,n\in\{0,1\}} \alpha_{m,n,\pm} \ket{m}\bra{n},
\end{equation}
where the coefficients are defined as
\begin{equation}
    \alpha_{m,n,\pm} = \sum_{i,j\in\{0,1\}} (\pm 1)^{i+j} \sqrt{p_i p_j} \braket{\gamma_{j,n} | \gamma_{i,m}}.
\end{equation}
Suppose Bob's measurement in X basis is defined as the POVM $\{\hat{M}_+,\hat{M}_-,\hat{M}_\bot\}$, where $\hat{M}_{\pm}$ and $\hat{M}_\bot$ corresponds to the result $\pm$ and inconclusive outcome, respectively. An essential assumption about Bob's POVMs is the $\hat{M}_\bot$ is the same in Z and X bases \cite{koashi2007RN398,tamaki2014RN69}. In this way, the joint probabilities $Y_{i,j}^{(z)}$ with $i,j = \pm$ in Eq. (\ref{eqa}) can be expressed as
\begin{equation}
    \label{eqb}
 Y_{i,j}^{(z)} = \frac{1}{2} \text{Tr} [\hat{M}_j \mathcal{E}(\varrho_i)].
\end{equation}
Here, the coefficient $1/2$ in Eq. (\ref{eqb}) is the probability that Alice obtains $\varrho_i$.

Below, we analyze how to estimate the phase error rate defined in Eq. (\ref{eqa}). According to Eq. (\ref{eqb}), the denominator of Eq. (\ref{eqa}) can be shown as
\begin{equation}
    \begin{aligned}
        \label{eqc}
        Y_{+,+}^{(z)} + Y_{+,-}^{(z)} + Y_{-,+}^{(z)} + Y_{-,-}^{(z)} 
        =& \frac{1}{2} \big\{\text{Tr} [(\hat{M}_+ + \hat{M}_-) \mathcal{E}(\varrho_+)] + \text{Tr} [(\hat{M}_+ + \hat{M}_-) \mathcal{E}(\varrho_-)] \big\} \\
        =& \frac{1}{2} \big\{\text{Tr} [(\mathbb{I} - \hat{M}_\bot) \mathcal{E}(\varrho_+)] + \text{Tr} [(\mathbb{I} - \hat{M}_\bot) \mathcal{E}(\varrho_-)] \big\} \\
        =& \text{Tr} [(\mathbb{I} - \hat{M}_\bot) \mathcal{E}(\rho_b)]\\
        =& s_1.
    \end{aligned}
\end{equation}
The last line in Eq. (\ref{eqc}) is due to the consistency of $\hat{M}_\bot$ in both bases. Hence, the denominator of Eq. (\ref{eqa}) is equal to the counting rate of the single-photon states $s_1$, which can be obtained directly in single-photon QKD. To obtain the numerator in Eq. (\ref{eqa}), we give the operator dominance condition that there exists $\beta_{i,\pm} \in \mathbb{R}$ satisfying the following formula
\begin{equation}
    \label{eqe}
    \varrho_\pm \leq \sum_{i\in\{0,1,+,-\}} \beta_{i,\pm} \rho_i.
\end{equation}
Further, we could restrict Eq. (\ref{eqe}) as equality in the majority of cases. If the prepared states $\rho_i,(i\in\{0,1,+,-\})$ form a tetrahedron on the Bloch sphere, then the right formula in Eq. (\ref{eqe}) can represent all of the Bloch vectors. If only three states are employed, the equality can still hold on condition that the states form a triangle on the Bloch sphere. We note that the condition can be easily satisfied in QKD systems. We take the phase-encoding scheme as an example to explain the validity of the equality of Eq. (\ref{eqe}). The phase-encoding state with phase $\theta$ can be expressed as $\ket{\omega_\theta} = (\ket{01} + e^{i\theta} \ket{10})/\sqrt{2}$, where $\ket{0}$ and $\ket{1}$ are the photon number states. The state can be re-expressed as $\ket{\omega_\theta} = \cos(\theta/2) \ket{0_z} + \sin(\theta/2) \ket{1_z}$ with the base transformation $\ket{0_z} = \ket{\omega_0}$ and $\ket{1_z} = -i\ket{\omega_\pi}$. Then the vector representation of the states $\rho_i$ are real-valued. Besides, the vector representation of $\varrho_\pm$ is real-valued if we select the purification satisfying $\braket{\gamma_{j,n} | \gamma_{i,m}} \in \mathbb{R}$. The equality of Eq. (\ref{eqe}) can be seen as a linear equation with the vector representation. The ranks of the coefficient and augmented matrices are both 3 and the linear equations have a unique solution $\beta_{i,\pm} \in \mathbb{R}$. These analyses and assumptions for the equality of Eq. (\ref{eqe}) are sufficient but unnecessary. We could directly solve Eq. (\ref{eqe}) directly. 

Based on Eq. (\ref{eqe}), we can obtain the upper bound of the probabilities in the numerator in Eq. (\ref{eqa}) as
\begin{equation}
    \begin{aligned}
    Y_{i,j}^{(z)} = \frac{1}{2} \text{Tr} [\hat{M}_j \mathcal{E}(\varrho_i)] \leq \sum_{k\in\{0,1,+,-\}} \frac{1}{2} \beta_{k,i} \text{Tr} [\hat{M}_j \mathcal{E}(\rho_k)] = \sum_{k\in\{0,1,+,-\}} \frac{1}{2} \beta_{k,i} Y_{k,j}.
    \end{aligned}
\end{equation}
Here, we define $Y_{k,j}$ with $k\in\{0,1,+,-\}$ and $j\in\{+,-\}$ as the conditional probability that Bob obtains the result $j$ when Alice prepares the states $\rho_i$ and Bob's measurement basis is X, which can be directly obtained in QKD systems. In this way, the upper bound of the phase error rate can be estimated as
\begin{equation}
 e_x = \sum_{k\in\{0,1,+,-\}} \frac{1}{2s_1} (\beta_{k,+} Y_{k,-} + \beta_{k,-} Y_{k,+}).
\end{equation}

In Eq. (\ref{eqd}), the state $\varrho_{\pm}$ is affected by the purification through the inner procuct $\braket{\gamma_{j,n} | \gamma_{i,m}}$. Hence the phase error rate may vary with different purification according to the probability in Eq. (\ref{eqb}). Since the purification is virtual and will not affect the actual process, we could optimize the purification freely to minimize the phase error rate. The only restriction of the purification is about the inner product $\braket{\gamma_{i,n} | \gamma_{i,m}} = \braket{m | \rho_i | n}$. We define a Gramian matrix of the (unnormalized) states $\ket{\gamma_{i,m}}$ as
\begin{equation}
 G = 
    \begin{bmatrix}
        \rho_0      & \sigma\\
        \sigma^\dag & \rho_1
    \end{bmatrix},
\end{equation}
where $\sigma$ is $2 \times 2$ and $\sigma_{m,n} = \braket{\gamma_{1,n} | \gamma_{0,m}}$. The Gramian matrix is positive, i.e., $G \geq 0$. Conversely, for any positive matrix, there exists a set of vectors such that their Gramian matrix corresponds to the given matrix. Hence we could only care about the Gramian matrix $G$ but not the specific states $\ket{\gamma_{i,m}}$. The optimization of the phase error rate can be simplified to semidefinite programming as $\min_{G \geq 0} e_x$, given the states $\rho_{i}$ and the probabilities $Y_{k,j}$ with $i,k\in\{0,1,+,-\}$ and $j\in\{+,-\}$.

In QKD systems, the probabilities $Y_{k,j}$ are obtained directly. The states $\rho_{i}$ should be characterized. As discussed above, the characterization might include inaccuracies. These discrepancies may lead to an underestimation of the phase error rate. To ensure security, we could analyze the upper bound of the phase error rate given the state spaces $V_i$ and conditional probabilities $Y_{k,j}$. We could calculate a presumptive phase error rate given any choice of the states $\rho_i \in V_i$. The maximum among all presumptive phase error rates gives an upper bound of the actual phase error rate. Therefore, the upper bound of the phase error rate is given by
\begin{equation}
    \label{eqm}
    \max_{\rho_i \in V_i} \min_{G \geq 0} e_x.
\end{equation}
With the upper bound on the phase error rate, we can determine the maximum information leakage to ensure the practical security of the protocol.

\subsection{Intensity and characterization imperfections}

In practical QKD systems, the weak coherent states are widely used as the physical carrier and the decoy-state method is employed to address the multi-photon issue. The secret keys are still distilled from the single-photon states. The SKR is given by \cite{gottesman2004RN266,lo2005RN81}
\begin{equation}
    \label{eqn}
 R_d = p_z^2 p_\mu \big\{ q_{1,\mu} s_1 [1 - h(e_x)] - \lambda_{\text{EC}} \big\},
\end{equation}
where $\mu$ is the intensity of the signal state for key generation, $p_\mu$ is the probability of preparing the signal state, and $\lambda_{\text{EC}}$ is the information revealed in the error correction step. In contrast to single-photon QKD, decoy-state QKD necessitates the estimation of both the phase error rate $e_x$ and the counting rate of single-photon states $s_1$. It is essential to characterize the encoding states and the distribution of the photon number states. Below we analyze how to estimate the parameters $s_1$ and $e_x$ with both encoding and intensity imperfections, as well as characterization imperfections.

Define $Y_{i,j}$, with $i\in\{0,1,+,-\}$ and $j\in\{0,1\}$, as the conditional probability that Bob obtains the result $j$ when Alice sends the single-photon state $\ket{1}\bra{1} \otimes \rho_i$ and Bob's measurement basis is Z. The counting rate of the single-photon state is defined as
\begin{equation}
    \label{eql}
 s_1 = \sum_{i,j\in\{0,1\}} p_i Y_{i,j}.
\end{equation}
Then the lower bound of $s_1$ can be obtained with the lower bounds of $Y_{i,j}$, $i,j\in\{0,1\}$. The phase error rate is defined as the same as that in single-photon QKD. To obtain the upper bound of the phase error rate, we should estimate the bounds of $Y_{i,j}$, $i\in\{0,1,+,-\}$ and $j\in\{+,-\}$. Overall, we need to estimate the bounds of $Y_{i,j}$ to obtain the bounds of $s_1$ and $e_x$. The conditional probability $Y_{i,j}$ can be expressed as
\begin{equation}
 Y_{i,j} = \text{Tr} [\hat{M}_j \mathcal{E} (\ket{1}\bra{1} \otimes \rho_i)].
\end{equation}
The operator dominance conditions for the estimation of the lower and upper bounds are given separately as 
\begin{equation}
    \begin{aligned}
        \label{eqf}
        \ket{1}\bra{1} &\geq \sum_{\tau\in\Pi} \beta_{\tau,0} \kappa_\tau,\\
        \ket{1}\bra{1} &\leq \sum_{\tau\in\Pi} \beta_{\tau,1} \kappa_\tau,
    \end{aligned}
\end{equation}
where $\beta_{\tau,0},\beta_{\tau,1} \in \mathbb{R}$. Once the conditions in Eq. (\ref{eqf}) are satisfied, we can obtain the bounds as
\begin{equation}
    \begin{aligned}
 Y_{i,j} &\geq \sum_{\tau\in\Pi} \beta_{\tau,0} Q_{i,j;\tau},\\
 Y_{i,j} &\leq \sum_{\tau\in\Pi} \beta_{\tau,1} Q_{i,j;\tau},
    \end{aligned}
\end{equation}
Here, $Q_{i,j;\tau} = \text{Tr} [\hat{M}_j \mathcal{E} (\kappa_\tau \otimes \rho_i)]$ is the counting rate that Bob obtains the result $j\in\{0,1\}$ ($\{+,-\}$) when Alice prepares the state coherent state with encoding $i$ and intensity $\tau$ and Bob measure in Z (X) basis. Below, we show how to construct the operator dominance conditions in Eq. (\ref{eqf}).

We consider the three-intensity decoy-state method, as an example, with intensity set $\Pi = \{\mu,\nu,o\}$, satisfying $0<o<\nu<\mu<1$. In practical QKD systems, the intensities of the coherent states may fluctuate due to the effects of limited modulation accuracy and environmental noise. Besides, the characterization of $\kappa_\tau$ might exhibit deviations from the actual states. Here we consider the state spaces $V_\tau$ for the actual state $\kappa_\tau \in V_\tau$. First, we define two operators as
\begin{equation}
    \begin{aligned}
        \label{eqg}
        \rho^\mu &= \frac{\kappa_\mu}{q_{0,\mu}} - \frac{\kappa_o}{q_{0,o}} = \sum_{s=1}^\infty \alpha_{s,\mu} \ket{s}\bra{s},\\
        \rho^\nu &= \frac{\kappa_\nu}{q_{0,\nu}} - \frac{\kappa_o}{q_{0,o}} = \sum_{s=1}^\infty \alpha_{s,\nu} \ket{s}\bra{s},
    \end{aligned}
\end{equation}
with coefficients defined as $\alpha_{s,\tau} = q_{s,\tau} / q_{0,\tau} - q_{s,o} / q_{0,o}$. By combining the two operators in Eq. (\ref{eqg}), we define the following operator
\begin{equation}
    \label{eqh}
    \alpha_{2,\nu} \rho^\mu - \alpha_{2,\mu} \rho^\nu = \sum_{s=1}^{\infty} \beta_s \ket{s}\bra{s},
\end{equation}
where $\beta_s = \alpha_{2,\nu} \alpha_{s,\mu} - \alpha_{2,\mu} \alpha_{s,\nu}$. To construct the operator dominance conditions in Eq. (\ref{eqf}), we need the following conditions
\begin{equation}
    \begin{aligned}
        \label{eqi}
        \alpha_{1,\nu} &> 0,\\
        \beta_1 &< 0.
    \end{aligned}
\end{equation}
These conditions can be easily satisfied and verified. For example, if $\mu > \nu > o$ for any $\mu\in\Pi_\mu$, $\nu\in\Pi_\nu$, we can guarantee these conditions. We could directly verify the conditions in Eq. (\ref{eqi}). Based on the mathematical induction method, we can conclude that $\alpha_{s,\nu} \geq 0$ and $\beta_s \geq 0$ with $s \geq 2$. The details are discussed in App. \ref{details}. According to Eqs. (\ref{eqg}), (\ref{eqh}) and (\ref{eqi}) and the corollaries above, we can obtain the operator dominance conditions for Eq. (\ref{eqf}) as $\ket{1}\bra{1} \leq \rho^\nu / \alpha_{1,\nu}$ and $\ket{1}\bra{1} \geq (\alpha_{2,\nu} \rho^\mu - \alpha_{2,\mu} \rho^\nu) / \beta_1$. Hence we give the coefficients for Eq. (\ref{eqf}) as
\begin{equation}
    \begin{aligned}
        \beta_{\mu,0} &= \frac{\alpha_{2,\nu}}{\beta_1 q_{0,\mu}},\\
        \beta_{\nu,0} &= -\frac{\alpha_{2,\mu}}{\beta_1 q_{0,\nu}},\\
        \beta_{o,0} &= \frac{\alpha_{2,\mu} - \alpha_{2,\nu}}{\beta_1 q_{0,o}},
    \end{aligned}
\end{equation}
and
\begin{equation}
    \begin{aligned}
        \beta_{\mu,1} &= 0,\\
        \beta_{\nu,1} &= \frac{1}{q_{0,\nu} \alpha_{1,\nu}},\\
        \beta_{o,1} &= \frac{1}{q_{0,o} \alpha_{1,\nu}}.
    \end{aligned}
\end{equation}
To address the imperfections of the characterization, we could estimate the lower and upper bounds of $Y_{i,j}$ as
\begin{equation}
    \begin{aligned}
        \label{eqk}
 Y_{i,j}^L &= \min_{V_\tau} \sum_{\tau\in\Pi} \beta_{\tau,0} Q_{i,j;\tau},\\
 Y_{i,j}^U &= \max_{V_\tau} \sum_{\tau\in\Pi} \beta_{\tau,1} Q_{i,j;\tau}.
    \end{aligned}
\end{equation}
With the bounds of $Y_{i,j}$ in Eq. (\ref{eqk}), we can calculate the lower bound of the counting rate for the single-photon states in Eq. (\ref{eql}) and the upper bound of the phase error rate in Eq. (\ref{eqm}). Then the SKR can be calculated using Eq. (\ref{eqn}). In this way, we address the imperfections in both state preparation and characterization in decoy-state QKD.

Our representation of the state preparation and characterization imperfections, along with the security analysis method, is readily applied to MDI-QKD protocol. Since MDI-QKD inherently immunes to the vulnerabilities at the detection side, it is beneficial to apply our method to MDI-QKD to enhance the overall security of QKD. Therefore, we provide a concise analysis of the applicability of our method to MDI-QKD in App. \ref{MDI_Imp}.

\begin{table}
    \caption{Gaussian parameters for the phase-encoding state preparation imperfections \cite{huang2023RN959}. $\bar{\theta}$ is the mean and $\sigma_\theta$ is the standard deviation.}
    \label{tab_phase}
    \centering
    \begin{tabular}{ccc}
    \toprule
 Phase    & $\bar{\theta}$ & $\sigma_\theta$ \\\hline
 0        & 0              & 0               \\
    $\pi/2$  & 0.4970$\pi$    & 0.0028$\pi$     \\
    $\pi$    & 0.9939$\pi$    & 0.0057$\pi$     \\
    $3\pi/2$ & 1.4910$\pi$    & 0.0084$\pi$ \\
    \bottomrule
    \end{tabular}
\end{table}

\begin{table}
    \caption{Gaussian parameters for the intensity state preparation imperfections \cite{huang2023RN959}. $\bar{\mu}$ is the mean and $\sigma_\mu$ is the standard deviation.}
    \label{tab_intensity}
    \centering
    \begin{tabular}{ccc}
        \toprule
 Intensity & $\bar{\mu}$ & $\sigma_\mu$ \\ \hline
 Vacuum    & $-$7.8E$-$3 & 0.0083       \\
 Decoy     & 0.236       & 0.0149       \\
 Signal    & 0.602       & 0.0258      \\ \bottomrule  
    \end{tabular}
\end{table}

\section{Simulation}

\begin{table}[]
    \caption{The difference between different cases in single-photon QKD.}
    \label{tab_case}
    \centering
\begin{tabular}{@{}lcc@{}}
\toprule
\multicolumn{1}{c}{}                            & State Preparation                 & State   Characterization \\ \midrule
\multicolumn{1}{c}{Ideal case} & Ideal                             & without deviation        \\
$\delta = 0$                   & $\bar{\theta}$ in Tab. \ref{tab_phase}             & without deviation        \\
$\delta > 0$                   & $\bar{\theta}$ in Tab. \ref{tab_phase}             & with deviation           \\
$\Delta = 0$                   & $\bar{\theta}, \sigma_\theta$ Tab. \ref{tab_phase} & without deviation        \\
$\Delta > 0$                   & $\bar{\theta}, \sigma_\theta$ Tab. \ref{tab_phase} & with deviation           \\ \bottomrule
\end{tabular}
\end{table}

\begin{figure}
    \centering
    \includegraphics[scale = 0.9]{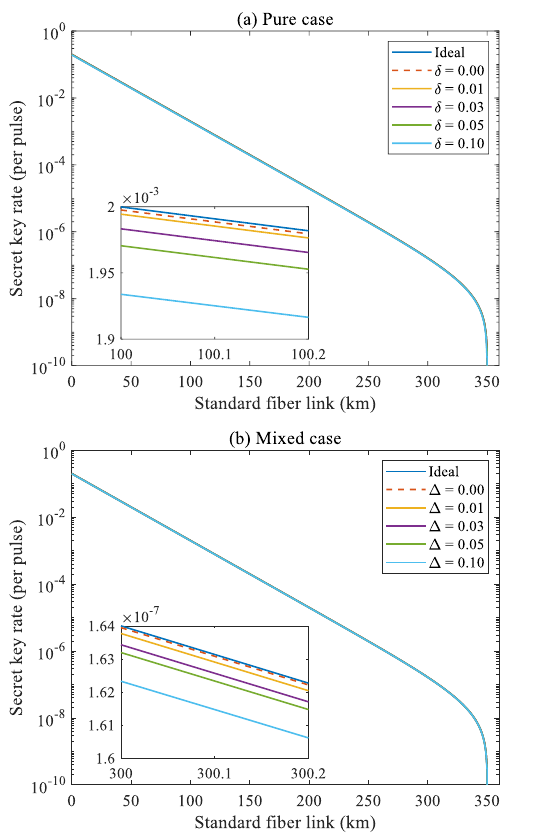}
    \caption{Secret key rate in logarithmic scale versus transmission distance between Alice and Bob in single-photon QKD with encoding and characterization imperfections. The subfigures illustrate the details at specific distances in linear scale. (a) The ideal case represents the scenario where both state encoding and state characterization are perfect. In other scenarios, the actual encoding states are still pure and set as the mean values in Tab. \ref{tab_phase}, but there exists a maximum state characterization deviation of $\delta$. (b) The ideal case represents the scenario where both state encoding and state characterization are perfect.  In other scenarios, the actual encoding states are mixed and the parameters are set according to Tab. \ref{tab_phase}. Besides, the maximum state characterization deviations of the mean and standard deviation are both set as $\Delta$.}
    \label{fig_sp_skr}
\end{figure}

To evaluate the performance of the method, we simulate the SKR of QKD in the presence of the state preparation and characterization imperfections. We adopt the parameters of the encoding and intensity imperfections from Ref. \cite{huang2023RN959}, which are measured in commercial QKD systems. These parameters are listed in Tabs. \ref{tab_phase} and \ref{tab_intensity}. Specifically, the parameters about phase $3\pi/2$ are not given in Ref. \cite{huang2023RN959} and are assumed based on other parameters of the other three phases in Tabs. \ref{tab_phase}. The means of phases 0, $\pi/2$, and $\pi$ deviate from their ideal values by 0, 0.003$\pi$, and 0.0061$\pi$, respectively. As the phase value increases, its mean deviates further from the ideal value. And we set the deviation for phase $3\pi/2$ as 0.009$\pi$. In the same way, we set the standard deviation of phase $3\pi/2$ as 0.0084$\pi$. Note that the parameters set for phase $3\pi/2$ are worse compared to other phases. Theoretically, the mean photon number is extremely close to but greater than zero. However, the fitted mean can occasionally become a very small negative number in Tab. \ref{tab_intensity} due to the following factors: the inherent electronic noise of detection system (e.g., oscilloscope, photodetector); the small offsets introduced in calibration and noise filtering procedures. This issue is precisely the problem that our paper aims to address: in practical QKD systems, the prepared quantum states suffer not only from state preparation imperfections, but also from state characterization imperfections. The negative mean intensity of the vacuum state is a concrete example of this characterization imperfection. We set the channel loss to the standard single-mode fiber parameter of 0.2 dB/km. The single-photon detector's detection efficiency is set to 80\%, and the dark count probability is set to 1E$-$8.

\begin{figure}
    \centering
    \includegraphics[scale = 0.9]{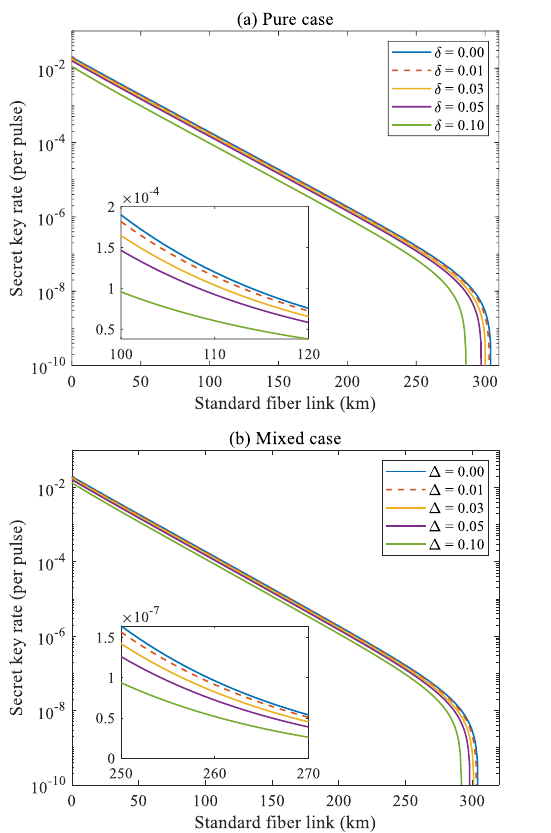}
    \caption{Secret key rate in logarithmic scale versus transmission distance between Alice and Bob in decoy-state QKD with intensity and characterization imperfections. The subfigures illustrate the details at specific distances in linear scale. The encoding is set to be ideal. (a) The intensities of the signal and decoy states are fixed as the mean values in Tab. \ref{tab_intensity} and the intensity of the vacuum state is set as 0.01. The maximum state characterization deviations are both set as $\delta'$. (b) The prepared states are set as mixed according to Tab. \ref{tab_intensity}. The maximum state characterization deviations of the mean and standard deviation are both set as $\Delta'$.}
    \label{fig_ds_skr}
\end{figure}

We first focus only on the encoding and characterization imperfections in single-photon QKD in Fig. \ref{fig_sp_skr}. We set the basis and bit selection probabilities as $p_z = p_x = 0.5$ and $p_0 = p_1 = 0.5$. In Fig. \ref{fig_sp_skr}(a), we consider the pure case where the modulated phases are fixed. The parameter $\delta$ characterizes the maximum state characterization deviation for all states. Suppose the characterized modulated phase is $\theta'$, we assume that the actual $\theta$ belongs to the neighborhood $U(\theta', \delta) = [\theta'-\delta,\theta'+\delta]$. Here, the neighborhood $U(\theta', \delta)$ characterizes the imperfection of characterization. We could use this to express the state spaces $V_i,i\in\{0,1,+,-\}$ as $\{ (\ket{01}+ e^{i \theta} \ket{10}) / \sqrt{2} | \theta \in U(\theta^\prime,\delta)\}$ so as to estimate the phase error rate in Eq. (\ref{eqm}). Since the state space is entirely determined by neighbourhood, an explicit and complex expression for the state spaces is not essential. We simulate the SKRs with different $\delta$ and compare them with the ideal case. The ideal case represents that both the encoding and the state characterization are perfect. And the case that $\delta = 0$ indicates that the encoding is imperfect while the state characterization remains perfect. The results show that the SKR under different conditions are nearly identical, even when the maximum state characterization deviation $\delta$ is as large as 0.1. The difference between different cases in single-photon QKD is summarize in Tab. \ref{tab_case}. Then we consider the mixed case in Fig. \ref{fig_sp_skr}(b). The parameter $\Delta$ characterizes the maximum state characterization deviation of the mean and standard deviation for all states. That is, the neighbourhoods $U(\theta', \Delta)$ and $U(\sigma_{\theta'}, \Delta)$ with measured $\theta'$ and $\sigma_{\theta'}$ depict the characterization deviation. With the neighbourhoods $U(\theta', \Delta)$ and $U(\sigma_{\theta'}, \Delta)$, while the explicit expressions of state spaces can be derived, they do not aid the analysis and are thus omitted for clarity. It can be seen that these SKRs with encoding imperfection and characterization deviations also nearly overlap with the ideal one in the mixed case. This verifies the tightness of the proposed method in the presence of encoding and characterization imperfections.

\begin{figure}
    \centering
    \includegraphics[scale = 0.64]{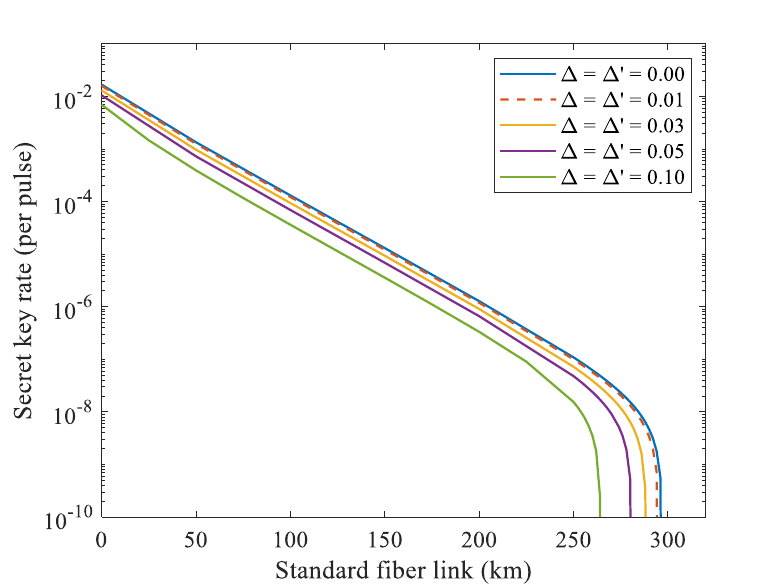}
    \caption{Secret key rate in logarithmic scale versus transmission distance between Alice and Bob in decoy-state QKD in the presence of imperfections in encoding, intensity, and characterization. The parameters of encoding and intensity are set according to Tabs. \ref{tab_phase} and \ref{tab_intensity}. The maximum state characterization deviations of the mean and standard deviation for encoding and intensity are set as $\Delta$ and $\Delta'$, respectively.}
    \label{fig_mixed_mixed}
\end{figure}

In Fig. \ref{fig_ds_skr}, we simulate the SKR in decoy-state QKD with intensity and characterization imperfections according to parameters in Tabs. \ref{tab_phase} and \ref{tab_intensity}. Here, since the physical intensity of light is always nonnegative, we adopt a truncated Gaussian distribution model for vacuum state: we truncate the Gaussian distribution at zero, considering only the portion of the distribution where the intensity is greater than or equal to zero, and renormalize the remaining probability density accordingly. This approach ensures that our analysis is physically meaningful and does not rely on unphysical negative intensities. Specifically, the truncated Gaussian distribution is given by
\begin{equation}
    f_T(x) = \frac{f(x)}{1 - F(0)},
\end{equation}
where $f(x)$ is the probability density function of the Gaussian distribution, adn $F(x)$ is the cumulative distribution function. The basis and bit selection probabilities are set the same as those in single-photon QKD. The probabilities for signal, decoy, and vacuum states are set as $p_\mu=$0.8, $p_\nu=$0.1, and $p_o=$0.1, respectively. We first consider the pure case in Fig. \ref{fig_ds_skr}(a). The parameter $\delta'$ characterizes the maximum state characterization deviation for all intensity states. Suppose the characterized intensity is $\mu'$, we assume that the actual $\mu$ belongs to the interval $[\mu'(1-\delta'), \mu'(1+\delta')]$. We note that it can be regarded as the ideal case when $\delta' = 0$ as our method does not assume the distribution $q_{j,\tau}$ in Eq. (\ref{eqr}). We simulate the SKRs with different $\delta'$. The results show that the SKR is compromised as the characterization deviation $\delta'$ increases. It can be seen that the impact of inaccuracies in intensity characterization on the secure key rate is greater than that of encoding characterization inaccuracies. A key reason is that the intensity characterization simultaneously determines the estimation of both the single-photon count rate and the phase error rate, while the encoding characterization only affects the estimation of the phase error rate. In Fig. \ref{fig_ds_skr}(b), the mixed case of intensity imperfection is considered by setting the maximum state characterization deviation $\Delta'$. Here, the intervals $[\mu'(1-\Delta'), \mu'(1+\Delta')]$ and $[\sigma_{\mu'}(1-\delta'), \sigma_{\mu'}(1+\delta')]$ with characterized $\mu'$ and $\sigma_{\mu'}$ depict the characterization deviation. Also, it corresponds to the ideal case when $\Delta' = 0$. The results obtained are similar to those under the pure state case.

In Fig. \ref{fig_mixed_mixed}, we consider the general case, where both encoding and intensity imperfections as well as characterization imperfections are simultaneously taken into account in the decoy-state QKD system. Here, we set $\Delta = \Delta'$ and simulate the SKR under various settings. The simulation results show that as the characterization deviation $\Delta$ and $\Delta'$ increase the SKR will decrease. The primary reason is the inaccuracy in characterizing the intensity imperfection. However, generally speaking, the characterization accuracy for the intensities can reach 0.01 or even lower in QKD systems. In this practical case, the SKR is almost identical to that with precise characterization. In this way, the effectiveness of the proposed method is verified despite the presence of imperfections in state preparation and characterization.

\begin{figure}
    \centering
    \includegraphics[scale = 0.57]{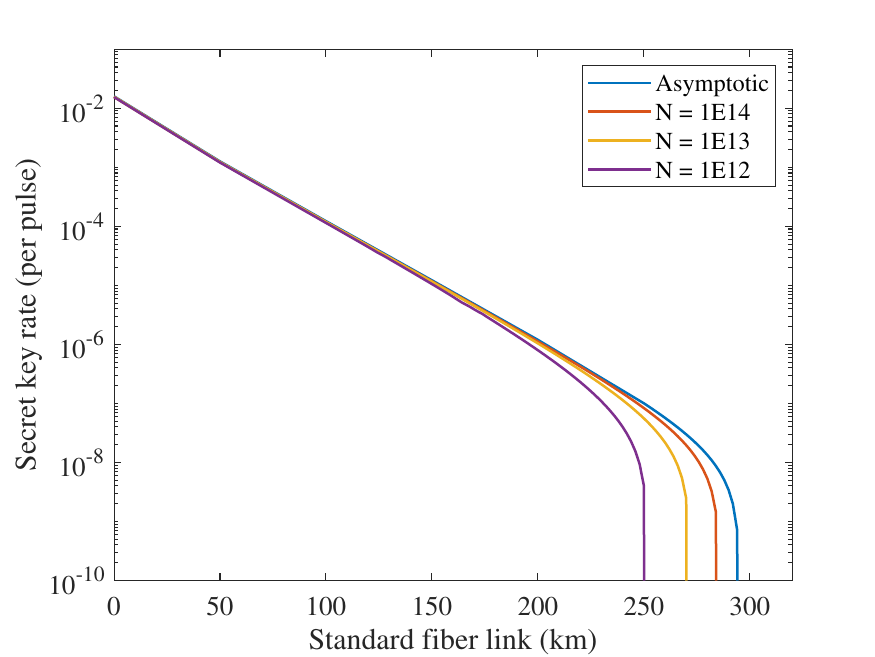}
    \caption{Secret key rate in logarithmic scale versus transmission distance between Alice and Bob in decoy-state QKD under practical conditions: finite-key effects, imperfections in encoding, intensity, and characterization. The parameters of encoding and intensity are set according to Tabs. \ref{tab_phase} and \ref{tab_intensity}. The maximum state characterization deviations of the mean and standard deviation for encoding and intensity are set as $\Delta = \Delta' = 0.01$.}
    \label{fig_mixed_finite}
\end{figure}

In Fig. \ref{fig_mixed_finite}, we consider the impact of finite-key effects in practical implementations. In realistic QKD systems, the total number of transmitted pulses is finite, which imposes limits on the achievable SKR. To evaluate the performance of our method under practical conditions, we simulate the SKR while incorporating imperfections in both state preparation and characterization, setting the deviation parameters as $\Delta = \Delta' = 0.01$. The results compare the SKR in both the asymptotic limit and under finite-key conditions with reasonable total pulse numbers $N=10^{12},10^{13}$, and $10^{14}$. The results show that the finite-key SKR gradually deviates from the asymptotic limit at longer distances, but this deviation becomes progressively weaker as the number of pulses $N$ increases. Notably, when $N=10^{12}$, the SKR remains nearly unaffected within a distance of 150 km and the maximum transmission distance can reach up to 250 km. When $N=10^{14}$, the finite-key SKR closely approaches the asymptotic limit up to 250 km, and the achievable secure distance extends to 284 km. These results demonstrate that our method remains effective in the finite-key regime, and suggest that the number of pulses can be optimized according to the target transmission distance to achieve near optimal performance.

\begin{figure}
    \centering
    \includegraphics[scale = 0.57]{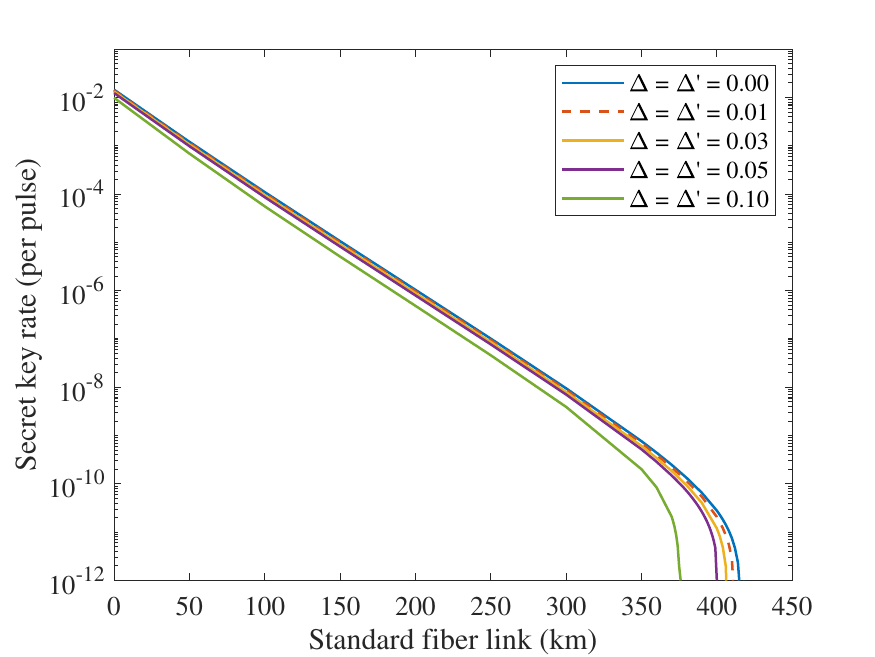}
    \caption{Secret key rate in logarithmic scale versus transmission distance between Alice and Bob in MDI-QKD in the presence of imperfections in encoding, intensity, and characterization. The parameters of encoding and intensity are set according to Tabs. \ref{tab_phase} and \ref{tab_intensity}.}
    \label{fig_mdi}
\end{figure}

In Fig. \ref{fig_mdi}, we simulate the SKR of MDI-QKD with imperfections in encoding, intensity, and characterization. Here, since only one set of parameters for the imperfections is available (\ref{tab_phase} and \ref{tab_intensity}), we adopt these parameters for both Alice and Bob for simplicity. However, the analysis remains applicable in general, as the security analysis method in App. \ref{MDI_Imp} does not rely on this specific assumption. The results show that the method remains highly applicable to MDI-QKD. Despite practical imperfections in encoding, intensity, and characterization, the SKR closely approaches the ideal case. Even with relatively large deviations ($\Delta = \Delta' = 0.1$), the method still enables MDI-QKD over distances exceeding 375 km.

\section{Discussion}

In QKD systems, the imperfections of state preparation can introduce vulnerabilities that compromise practical security. The imperfections in state characterization further exacerbate the challenges by introducing additional inaccuracies. In this work, we present the practical security analysis of QKD in practical scenario involving imperfections of encoding, intensity and the state characterization. To accommodate these imperfections simultaneously, we provide a unified framework based on the operator dominance method and give a tight estimation of the SKR. Our method is based on a characterization of the quantum states, while allowing for some tolerance in the accuracy of the state characterization. This flexibility helps reduce conservativeness in parameter estimation so as to achieve high QKD performance while maintaining practicality. Moreover, we offer a concise analysis to prove the suitability to MDI-QKD, enabling practical security on both Alice’s and Bob’s sides. These advancements contribute to the development of more reliable and secure quantum networks.

Future advances in practical QKD security could be realized by extending security analyses to include additional imperfections, including side-channel effects and correlations. First, the present analysis requires the qubit assumption, which inherently excludes potential side channels. However, in practice, the encoding and intensity modulation processes may reflect inadvertently in other degrees of freedom \cite{pereira2019RN65}, creating exploitable side channels. The effects of side channels cannot be directly characterized using the operator dominance method, especially since the relevant side-channel parameters are often inaccessible or poorly characterized in realistic implementations. Second, the above analyses do not incorporate the corrections between consecutively prepared quantum states. Such corrections undermine the independence of quantum states, which can be essentially regarded as a form of information leakage \cite{guillermo2023RN1332}. Fortunately, several promising approaches have been proposed to address these imperfections \cite{yoshino2018RN71,pereira2019RN65,mizutani2019RN61,pereira2020RN255,zapatero2021RN666,sixto2022RN886,pereira2023RN952,guillermo2023RN1332,2024RN1234,pereira2025RN1330,gnanapandithan2025RN1303,marcomini2025RN1329,li2025RN1342}. Therefore, future efforts could focus on enhancing our framework by integrating these advanced techniques and explicitly addressing practical experimental imperfections. Overcoming these challenges will be essential for the deployment of secure, high-performance, and scalable quantum communication networks.

\section*{Acknowledgments}

This work was supported by Henan Science and Technology Major Project of the Department of Science \& Technology of Henan Province (No. 241100210400), National Key Research and Development Program of China (No. 2020YFA0309702), National Natural Science Foundation of China (Nos. U2130205, 62101597), Natural Science Foundation of Henan (No. 242300421219).

\appendix

\section{MDI-QKD with imperfections in state preparation and characterization}
\label{MDI_Imp}

We first consider the encoding and characterization imperfections in single-photon MDI-QKD. We suppose the prepared states at Alice and Bob's sides both can be denoted as $\rho_i$, ($i\in\{0,1,+,-\}$) for simplicity. But the subsequent analysis remains valid even when the states prepared by Alice and Bob are different. In the entanglement version, we suppose both Alice and Bob prepare the entanglement states $\ket{\Phi_z}_{aAc_1}$ and $\ket{\Phi_z}_{bBc_2}$ in Eq. (\ref{eqo}). Then keep the systems $aA$ and $bB$ locally, and send the systems $c_1$ and $c_2$ to an untrusted party (named Charlie). Charlie is supposed to perform interference measurement and announce the results $\{0,1,\bot\}$, where 0 and 1 correspond to the announcement of the Bell state $(\ket{01} \pm \ket{10})/\sqrt{2}$, and $\bot$ corresponds to other invalid measurements. After the announcement, the states corresponding to the announced results $k \in \{0,1\}$ can be expressed as
\begin{equation}
    \begin{aligned}
        \rho_{(k)} = \frac{1}{p_{(k)}} \hat{C}_k (\rho_{ac_1} \otimes \rho_{bc_2}) \hat{C}_k^\dagger,
    \end{aligned}
\end{equation}
where $\hat{C}_k$ represents Charlie's operation when announcing result $k$, and
\begin{equation}
 p_{(k)} = \text{Tr}[\hat{C}_k (\rho_{ac_1} \otimes \rho_{ac_2}) \hat{C}_k^\dagger],
\end{equation}
is the probability of announcing result $k$, and the states are defined as
\begin{equation}
    \begin{aligned}
        \rho_{ac_1} &= \text{Tr}_{A} [\hat{P}(\ket{\Phi_z}_{aAc_1})],\\
        \rho_{bc_2} &= \text{Tr}_{B} [\hat{P}(\ket{\Phi_z}_{bBc_2})].
    \end{aligned}
\end{equation}
In this way, we can define the phase error rate corresponding to result $k \in \{0,1\}$ below similarly as that in Eq. (\ref{eqa})
\begin{equation}
    \begin{aligned}
        \label{eqp}
 e_{x_0} &= \frac{Y_{+,-}^{(z),0} + Y_{-,+}^{(z),0}}{Y_{+,+}^{(z),0} + Y_{+,-}^{(z),0} + Y_{-,+}^{(z),0} + Y_{-,-}^{(z),0}} = \frac{1}{p_{(0)}} (Y_{+,-}^{(z),0} + Y_{-,+}^{(z),0}),\\
 e_{x_1} &= \frac{Y_{+,+}^{(z),1} + Y_{-,-}^{(z),1}}{Y_{+,+}^{(z),1} + Y_{+,-}^{(z),1} + Y_{-,+}^{(z),1} + Y_{-,-}^{(z),1}} = \frac{1}{p_{(1)}} (Y_{+,+}^{(z),1} + Y_{-,-}^{(z),1}),
    \end{aligned}
\end{equation}
where $Y_{i,j}^{(z),k}$, with $i,j=\pm$ and $k\in\{0,1\}$, is the joint probability that Alice and Bob obtain results $i$ and $j$ in X basis and Charlie announces result $k$. We can define the overall phase error rate using the convexity of the binary entropy function as
\begin{equation}
 e_x^{\text{MDI}} = \frac{p_{(0)} e_{x_0} + p_{(1)} e_{x_1}}{p_{(0)} + p_{(1)}}.
\end{equation}
Note that Alice and Bob's measurements are performed on the systems $a$ and $b$, which are independent of Charlie's measurement. Hence, the probability $Y_{i,j}^{(z),k}$ can be expressed as
\begin{equation}
    \begin{aligned}
 Y_{i,j}^{(z),k} &= p_{(k)} \text{Tr} \big[ \ket{ij}\bra{ij} \rho_{(k)} \big] \\
        &= \text{Tr} \big[\ket{ij}\bra{ij} \otimes \hat{C}_k^\dagger \hat{C}_k (\rho_{ac_1} \otimes \rho_{bc_2}) \big] \\
        &= \frac{1}{4} \text{Tr} \big[\hat{C}_k^\dagger \hat{C}_k (\varrho_i \otimes \varrho_j) \big],
    \end{aligned}
\end{equation}
where the states $\varrho_i$, with $i=\pm$, are defined in Eq. (\ref{eqd}). Then with the operator dominance condition in Eq. (\ref{eqe}), we can obtain
\begin{equation}
    \varrho_i \otimes \varrho_j \leq \sum_{m,n\in\{0,1,+,-\}} \beta_{m,i} \beta_{n,j} \rho_m \otimes \rho_n.
\end{equation}
In this way, we could estimate the upper bound of $Y_{i,j}^{(z),k}$ as 
\begin{equation}
    \begin{aligned}
 Y_{i,j}^{(z),k} &\leq \frac{1}{4} \sum_{m,n\in\{0,1,+,-\}} \beta_{m,i} \beta_{n,j} \text{Tr} \big[\hat{C}_k^\dagger \hat{C}_k (\rho_m \otimes \rho_n) \big] \\
        &= \frac{1}{4} \sum_{m,n\in\{0,1,+,-\}} \beta_{m,i} \beta_{n,j} Y_{m,n}^{(k)},
    \end{aligned}
\end{equation}
where we have defined
\begin{equation}
    \label{eqq}
 Y_{m,n}^{(k)} = \text{Tr} \big[\hat{C}_k^\dagger \hat{C}_k (\rho_m \otimes \rho_n) \big],
\end{equation}
which is the probability of obtaining Charlie's announced result $k$ when Alice and Bob send the states $\rho_m$ and $\rho_n$, respectively. Note that the probability $Y_{m,n}^{(k)}$ can be directly obtained. The optimal of the phase error rate is the same as that in Eq. (\ref{eqm}). In this way, we present the parameter estimation method in single-photon MDI-QKD. The SKR of single-photon MDI-QKD is shown as \cite{lo2012RN72,ma2012RN73}
\begin{equation}
 R_{s}^{\text{MDI}} = p_z^2 s_{11} [1 - h(e_x^{\text{MDI}}) - h(e_z)],
\end{equation}
where $p_z$ is the probability of selecting Z basis, $s_{11}$ is the counting rate in Z basis (i.e., the counting rate when Alice and Bob send the single-photon states), and $e_z$ is its bit error rate.

In decoy-state MDI-QKD, the key is to estimate the bounds of $Y_{m,n}^{(k)}$ defined in Eq. (\ref{eqq}) corresponding to the single-photon states. The counting rate of the single-photon states is shown as
\begin{equation}
 s_1^{\text{MDI}} = \sum_{m,n,k\in{0,1}} p_{(k)} p_i p_j Y_{m,n}^{(k)}.
\end{equation}
And the phase error rate can be estimated with the bounds of $Y_{m,n}^{(k)}$ the same as that in single-photon MDI-QKD. Below we show how to estimate the bounds of $Y_{m,n}^{(k)}$, which can be re-expressed by introducing the photon number as
\begin{equation}
 Y_{m,n}^{(k)} = \text{Tr} \big[\hat{C}_k^\dagger \hat{C}_k (\ket{1}\bra{1} \otimes \rho_m \otimes \ket{1}\bra{1} \otimes  \rho_n) \big].
\end{equation}
With the operator dominance conditions in Eq. (\ref{eqf}), we can derive the bounds as
\begin{equation}
    \begin{aligned}
 Y_{m,n}^{(k),L} &= \min_{V_{\tau_a},V_{\tau_b}} \sum_{\tau_a,\tau_b\in\Pi} \beta_{\tau_a,0} \beta_{\tau_a,0} Q_{m,n;\tau_a,\tau_b}^{(k)},\\
 Y_{m,n}^{(k),U} &= \max_{V_{\tau_a},V_{\tau_b}} \sum_{\tau_a,\tau_b\in\Pi} \beta_{\tau_a,1} \beta_{\tau_a,1} Q_{m,n;\tau_a,\tau_b}^{(k)},\\
    \end{aligned}
\end{equation}
where $Q_{m,n;\tau_a,\tau_b}^{(k)}$ is the probability that Alice and Bob's encoding bits are $m$ and $n$, the intensities are $\tau_a$ and $\tau_b$, and Charlie announces result $k$. Then we can calculate the lower bound of the counting rate and the upper bound of the phase error rate for the single-photon states. The SKR of the decoy-state MDI-QKD is given as \cite{lo2012RN72,ma2012RN73}
\begin{equation}
 R_{d}^{\text{MDI}} = p_z^2 p_\mu^2 \big\{ q_{1,\mu}^2 s_{11}^{\text{MDI}} [1 - h(e_x^{\text{MDI}})] - \lambda_{\text{EC}} \big\},
\end{equation}
where $p_\mu$ is the probability of preparing the signal states, $q_{1,\mu}$ is the probability of the single-photon states as shown in Eq. (\ref{eqr}).

\section{Details for operator dominance conditions}
\label{details}

The conditions in Eq. (\ref{eqi}) can be verified easily, as they only involve simple numerical calculations. Below, we provide details of obtaining the corollaries that $\alpha_{s,\nu} \geq 0$ and $\beta_s \geq 0$ with $s \geq 2$. 

For simplicity, we first define the formulas as
\begin{equation}
 C_{\tau_0,\tau_1;s} = \frac{\tau_0^s - \tau_1^s}{\tau_0 - \tau_1} = \sum_{n=0}^{s-1} \tau_0^n \tau_0^{s-n-1},
\end{equation}
and that
\begin{equation}
    \begin{aligned}
 C_{\tau_0,\tau_1;s}^{\min} &= \min_{\tau_0\in\Pi_{\tau_0},\tau_1\in\Pi_{\tau_1}} C_{\tau_0,\tau_1;s},\\
 C_{\tau_0,\tau_1;s}^{\max} &= \max_{\tau_0\in\Pi_{\tau_0},\tau_1\in\Pi_{\tau_1}} C_{\tau_0,\tau_1;s}.\\
    \end{aligned}
\end{equation}
We can verify the following formula with $s \geq 2$, where the range of the summation parameters is omitted for simplicity,
\begin{equation}
    \begin{aligned}
        \label{eqs}
        \alpha_{s,\nu} &= \frac{q_{s,\nu}}{q_{0,\nu}} - \frac{q_{s,o}}{q_{0,o}} = \frac{q_{s,\nu} q_{0,o} - q_{0,\nu} q_{s,o}}{q_{0,\nu} q_{0,o}} \\
        &= \sum_{\nu_0\in\Pi_\nu,0_0\in\Pi_o} \frac{f_{\nu_0} f_{o_0}}{q_{0,\nu} q_{0,o}} (p_{s,\nu_0} p_{0,o_0} - p_{0,\nu_0} p_{s,o_0})\\
        &= \sum_{\nu_0\in\Pi_\nu,0_0\in\Pi_o} \frac{f_{\nu_0} f_{o_0}}{q_{0,\nu} q_{0,o}} \frac{\nu_0^s - o_0^s}{e^{\nu_0+o_0}s!} \\
        &= \sum_{\nu_0\in\Pi_\nu,0_0\in\Pi_o} \frac{f_{\nu_0} f_{o_0}}{q_{0,\nu} q_{0,o}} \frac{(\nu_0 - o_0)C_{\nu_0,o_0;s}}{e^{\nu_0+o_0}s!} \\
        &\geq \frac{C_{\nu_0,o_0;s}^{\min} (s-1)!}{C_{\nu_0,o_0;s-1}^{\max} s!} \sum_{\nu_0,o_0} \frac{f_{\nu_0} f_{o_0}}{q_{0,\nu} q_{0,o}} \frac{(\nu_0 - o_0)C_{\nu_0,o_0;s-1}}{e^{\nu_0+o_0}(s-1)!} \\
        &= \frac{C_{\nu_0,o_0;s}^{\min} (s-1)!}{C_{\nu_0,o_0;s-1}^{\max} s!} \sum_{\nu_0,o_0} \frac{f_{\nu_0} f_{o_0}}{q_{0,\nu} q_{0,o}} \frac{\nu_0^{s-1} - o_0^{s-1}}{e^{\nu_0+o_0}(s-1)!} \\
        &= \frac{C_{\nu_0,o_0;s}^{\min} (s-1)!}{C_{\nu_0,o_0;s-1}^{\max} s!} \alpha_{s-1,\nu}, \\
    \end{aligned}
\end{equation}
Here, we define the $f_{\tau_0}$ as the probability that a state intended to have intensity $\tau$ is actually prepared with an intensity $\tau_0$. This arises from the imperfections in the state preparation. As the fraction in the last line of Eq. (\ref{eqs}) is greater than or equal to 0, the sign of $\alpha_{s,\nu}$ is the same as that of $\alpha_{s-1,\nu}$, or $\alpha_{s,\nu}$ is equal to 0. Therefore, if $\alpha_{1,\nu} > 0$, we can conclude that $\alpha_{s,\nu} \geq 0$ with $s\geq 2$.

It is evident that $\beta_2 = \alpha_{2,\nu} \alpha_{2,\mu} - \alpha_{2,\mu} \alpha_{2,\nu} = 0$. We give the following formula with $s \geq 3$
\begin{equation}
    \begin{aligned}
        \beta_s &=  \alpha_{2,\nu} \alpha_{s,\mu} - \alpha_{2,\mu} \alpha_{s,\nu} \\
        &= \bigg(\frac{q_{2,\nu}}{q_{0,\nu}} - \frac{q_{2,o}}{q_{0,o}}\bigg) \bigg(\frac{q_{s,\mu}}{q_{0,\mu}} - \frac{q_{s,o}}{q_{0,o}}\bigg) \\
        &- \bigg(\frac{q_{2,\mu}}{q_{0,\mu}} - \frac{q_{2,o}}{q_{0,o}}\bigg) \bigg(\frac{q_{s,\nu}}{q_{0,\nu}} - \frac{q_{s,o}}{q_{0,o}}\bigg) \\
        &= \sum_{\mu_0,\nu_0,o_0,o_1} \frac{f_{\mu_0}f_{\nu_0}f_{o_0}f_{o_1}}{q_{0,\nu} q_{0,\mu} q_{0,o}^2} \frac{D(\mu_0,\nu_0,o_0,o_1;s)}{e^{\mu_0+\nu_0+o_0+o_1}2!j!}.
    \end{aligned}
\end{equation}
Here, we have defined the parameter
\begin{equation}
    \begin{aligned}
        &D(\mu_0,\nu_0,o_0,o_1;s)\\
        \triangleq & (\nu_0^2 - o_0^2)(\mu_0^s - o_1^s) - (\mu_0^2 - o_0^2)(\nu_0^s - o_1^s)\\
 = & (\nu_0^2 - o_0^2)(\mu_0^{s-1} - o_1^{s-1}) \frac{\mu_0^s - o_1^s}{\mu_0^{s-1} - o_1^{s-1}} \\
        &- (\mu_0^2 - o_0^2)(\nu_0^{s-1} - o_1^{s-1}) \frac{\nu_0^s - o_1^s}{\nu_0^{s-1} - o_1^{s-1}} \\
        \geq & C^{\min} [(\nu_0^2 - o_0^2)(\mu_0^{s-1} - o_1^{s-1}) - (\mu_0^2 - o_0^2)(\nu_0^{s-1} - o_1^{s-1})] \\
 =& C^{\min} D(\mu_0,\nu_0,o_0,o_1;s-1),
    \end{aligned}    
\end{equation}
with
\begin{equation}
 C^{\min}_s = \min_{\mu_0,\nu_0,o_1} \bigg\{ \frac{\mu_0^s - o_1^s}{\mu_0^{s-1} - o_1^{s-1}}, \frac{\nu_0^s - o_1^s}{\nu_0^{s-1} - o_1^{s-1}} \bigg\} \geq 0.
\end{equation}
Therefore, based on the mathematical induction method, we can obtain that
\begin{equation}
    \beta_s \geq C^{\min}_s \beta_{s-1} \geq 0, s \geq 2.
\end{equation}

\bibliographystyle{spphys}
\bibliography{Ref}

\end{document}